\documentclass[a4paper, oneside, twocolumn, notitlepage, 10pt]{extarticle_ecoc2015}
\usepackage{ecoc2015}
\usepackage{dblfloatfix}

\usepackage{pgfplots}
\usetikzlibrary{backgrounds,intersections,positioning}
\usetikzlibrary{external}
\usepackage{bm}
\usepackage{alucolor}
\usepackage{comment}
\usepackage{acronym}

\newcommand\blfootnote[1]{%
  \begingroup
  \renewcommand\thefootnote{}\footnote{#1}%
  \addtocounter{footnote}{-1}%
  \endgroup
}
\colorlet{figyellow}{yellow!40!white}
\colorlet{figred}{red!50!white}
\colorlet{figblue}{blue!20!white}
\colorlet{figgreen}{green!30!white}
\colorlet{figgray}{black!10!white}

\newcommand{\msR}{\mathsf{R}}

\newcommand{\Es}{\mathsf{E}_{\mathsf{s}}}
\newcommand{\No}{\mathsf{N}_0}

\newcommand{\GMI}{\mathsf{GMI}}

\acrodef{ADC}{analog-to-digital converter}
\acrodef{AIR}{achievable information rate}
\acrodef{ASE}{amplified spontaneous emission}
\acrodef{ASIC}{application specific integrated circuit}
\acrodef{AWGN}{additive white Gaussian noise}
\acrodef{BCH}{Bose-Chaudhuri-Hocquenghem}
\acrodef{BDD}{bounded distance decoding}
\acrodef{BEC}{binary erasure channel}
\acrodef{BER}{bit error rate}
\acrodef{BICM}{bit-interleaved coded modulation}
\acrodef{BMD}{bit metric decoding}
\acrodef{BP}{belief propagation}
\acrodef{BPSK}{binary phase shift keying}
\acrodef{BSC}{binary symmetric channel}
\acrodef{BTC}{block turbo code}
\acrodef{CM}{coded modulation}
\acrodef{CN}{check node}
\acrodef{CRC}{cyclic redundancy check}
\acrodef{DAC}{digital-to-analog converter}
\acrodef{DD}{direct detection}
\acrodef{DMC}{discrete memoryless channel}
\acrodef{DMSC}{discrete memoryless symmetric}
\acrodef{DSP}{digital signal processing}
\acrodef{EDFA}{Erbium-doped fiber amplifier}
\acrodef{EM}{expectation-maximization}
\acrodef{EXIT}{extrinsic information transfer}
\acrodef{FEC}{forward error correction}
\acrodef{FER}{frame error rate}
\acrodef{FPGA}{field-programmable gate array}
\acrodef{GMI}{generalized mutual information}
\acrodef{GMM}{Gaussian mixture model}
\acrodef{GPU}{graphical processing unit}
\acrodef{HDD}{hard decision decoding}
\acrodef{HD}{hard decision}
\acrodef{IID}{independent and identically distributed}
\acrodef{KDE}{kernel density estimator}
\acrodef{LLR}{log-likelihood ratio}
\acrodef{LDPC}{low-density parity-check}
\acrodef{MAP}{maximum a posteriori}
\acrodef{MET}{multi-edge type}
\acrodef{MI}{mutual information}
\acrodef{ML}{maximum likelihood}
\acrodef{MLC}{multi-level coding}
\acrodef{MSD}{multi-stage decoding}
\acrodef{NB}{nonbinary}
\acrodef{NCG}{net coding gain}
\acrodef{OOK}{on-off keying}
\acrodef{OSNR}{optical signal-to-noise ratio}
\acrodef{OTN}{optical transport network}
\acrodef{PAM}{pulse amplitude modulation}
\acrodef{PAS}{probabilistic amplitude shaping}
\acrodef{PC}{product code}
\acrodef{PDF}{probability density function}
\acrodef{PEG}{progressive edge growth}
\acrodef{PM}{polarization-multiplexed}
\acrodef{PON}{passive optical network}
\acrodef{PRBS}{pseudo-random binary sequence}
\acrodef{QAM}{quadrature amplitude modulation}
\acrodef{QC}{Quasi-Cyclic}
\acrodef{QPSK}{quadrature phase shift keying}
\acrodef{RF}{radio frequency}
\acrodef{RS}{Reed-Solomon}
\acrodef{SC}{spatially coupled}
\acrodef{SCC}{serially concatenated code}
\acrodef{SD}{soft-decision}
\acrodef{SDD}{soft decision decoding}
\acrodef{SER}{symbol error rate}
\acrodef{SNR}{signal-to-noise ratio}
\acrodef{SOP}{state of polarization}
\acrodef{SSMF}{standard single-mode fiber}
\acrodef{TPC}{turbo product code}
\acrodef{VN}{variable node}
\acrodef{VLSI}{very large scale integration}
\acrodef{WDM}{wavelength division multiplex}
\acrodef{WSS}{wavelength selective switch}

\begin{document}
\selectlanguage{american}    


\title{Performance Metrics for Communication Systems\\ with Forward Error Correction\vspace*{-1ex}}%


\author{
    Laurent Schmalen\vspace*{-1ex}}

\maketitle                  


\begin{strip}
 \begin{author_descr}

   Nokia Bell Labs, Lorenzstr. 10, Stuttgart, Germany,
   \uline{\{firstname.lastname\}@nokia-bell-labs.com}
 \end{author_descr}
\end{strip}


\begin{strip}
  \begin{ecoc_abstract}
We revisit performance metrics for optical communication systems with FEC. We illustrate the concept of universality and discuss the most widespread performance thresholds. Finally, we show by example how to include FEC into transmission experiments.\vspace*{-1ex}
  \end{ecoc_abstract}
\end{strip}


\section{Introduction}
The design of many optical communication systems requires the heavy use of transmission experiments to verify models, assumptions and complete systems. This is mostly due to the absence of a rigorous and widely accepted channel model that includes all the effects and impairments of transceivers and optical fibers. The fiber-optical communication channel is however rather static, enabling the relatively easy reproducibility of the experiments.\blfootnote{This work is supported by the German BMBF in the scope of the CELTIC+ project SENDATE-TANDEM.}

With the advent of coherent optical communications and in particular \ac{SDD}\cite{schmalen2013bltj}, the widely used pre-FEC \ac{BER} threshold was no longer an acceptable metric. This was first realized in\cite{LevenMI} for early coherent system and performance metrics based on information theory were proposed. In\cite{Alvarado2015b_JLT}, this approach was extended to \ac{BICM},  the \textit{de-facto} \ac{CM} standard for today's systems. It has now become customary to characterize a \ac{FEC} code by an achievable rate and output BER, where the achievable rate depends on the \ac{CM} scheme used. In transmission experiments, an estimate of the achievable rate is computed and used to evaluate if the BER can be achieved or not. This assumes (often without stating so) that the utilized \ac{FEC} code is \textit{universal}.

In this paper, we will first introduce the concept of \textit{universality} and then give some hints on the achievable rate to be used in various circumstances. We wrap the paper by illustrating how to include FEC into transmission experiments.\vspace*{-1ex}

\newcommand{\bx}{\bm{x}}

\section{Threshold-based FEC Performance Prediction}\label{sec:s5_threshold}
While thresholds are a perfectly fine tool for predicting the performance of some \ac{FEC} schemes with \ac{HDD}, the use of \ac{FEC} schemes with \ac{SDD} together with varying modulation formats and transmission links requires more caution. This is due to the fact that \ac{SDD} is a statistical estimation process that critically relies on the knowledge of the channel model. In this scenario, the concept of \ac{FEC} universality is crucial when using \textit{thresholds}. 

A pair of \ac{FEC} code and its decoder are said to be \textit{universal}, if the performance of the code (in terms of post-FEC \ac{BER}) does not depend on the channel (whole chain between the \ac{FEC} encoder output $\bx$ and the decoder \ac{LLR} input $\bm{l}$, including modulation/demodulation, DSP, ADC and DAC, fiber transmission, filtering and amplification including noise), provided that the \ac{CM} scheme and the achievable rate (e.g., the GMI) are fixed and identical. 

\begin{figure}[b!]
\vspace*{-3ex}
\includegraphics[width=\columnwidth]{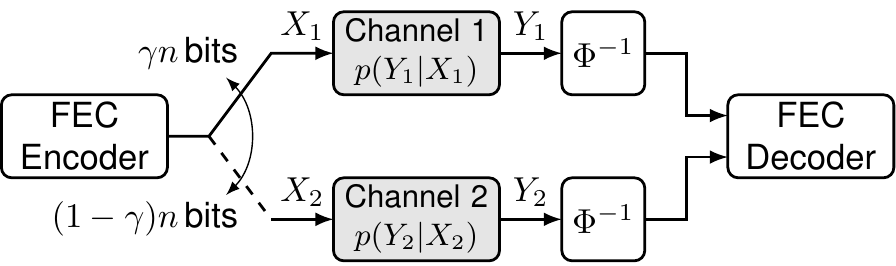}
\caption{Definition of universality of \acs{FEC} schemes according to\cite{sanaei2008design}. Channels 1 and 2 have the same GMI (achievable rate).}
\label{fig:universalitydef}
\end{figure}

Many practical \ac{LDPC} codes are conjectured to be approximately universal\cite{Franceschini06}. Polar codes are examples of \textit{non}-universal codes that need to be redesigned for every different channel. We can define universality\cite{sanaei2008design} using Fig.~\ref{fig:universalitydef}. We assume \ac{BICM} and bit-wise \ac{SDD}. An \ac{FEC} encoder generates a codeword of $n$ bits. We transmit these bits over 2 channels with different (memoryless) channel transition \acp{PDF}: Channel 1 with \ac{PDF} $p(y_1|x_1)$ and channel 2 with \ac{PDF} $p(y_2|x_2)$. Both channels have \textit{identical} $\GMI$. A fraction $\gamma n$ of the bits is transmitted over channel 1, while the remaining bits are transmitted over channel 2. At the receiver, \ac{BMD} is carried out ($\Phi^{-1}$) and the combined sequence is decoder. A code is \textit{universal} for channels 1 and 2 if the post-\ac{FEC} \ac{BER} is \textit{independent} of $\gamma$.

\newcommand{\argmin}{\mathop{\mathrm{arg\,min}}}

\begin{figure}[t!]
\centering
\includegraphics[width=\columnwidth]{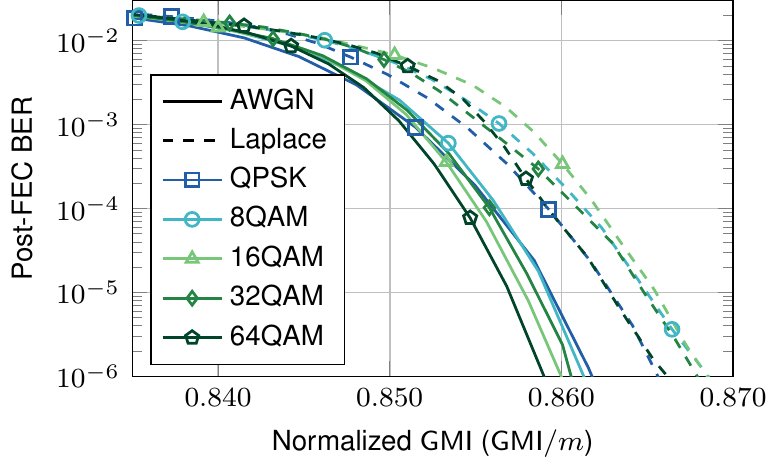}
\caption{Illustration of the non-universality for \ac{BICM}.}
\label{fig:modu_universality}
\end{figure}

In\cite{schmalen2017performance}, the impact of strong quantization of the channel output on universality was shown for  \ac{CM} with nonbinary \ac{LDPC} codes: quantization led to significantly different performance for constant achievable rate (MI). We illustrative the concept of non-universality of a common \ac{FEC} decoder with \ac{BICM}. We consider a regular QC-\ac{LDPC} code of rate $R=0.8$  with \ac{BICM} and 5 different constellations: QPSK, 8QAM ($\mathcal{C}_3$ of~Fig.~3 in\cite{schmalen2017performance}), 16QAM, 32QAM, and 64QAM. Let $m$ denote the number of bits mapped to each 2D symbol. We consider transmission over both AWGN and the Laplace channel, which adds noise (per dimension) with\vspace*{-1.5ex} 
\[
p_{Y|X}(y|x) = \frac{1}{2b}\exp\left(-\frac{|y-x|}{b}\right),\ b=\frac{\left(\Es/\No\right)^{-\frac{1}{2}}}{2}\,.
\]
At the receiver, we employ, after \ac{BMD}, scaled min-sum decoding with 10 iterations. All setups are compared at the same normalized $\GMI/m$. The post-FEC BER results are shown in Fig.~\ref{fig:modu_universality}. The GMI is only an approximately good threshold of the performance. If the channel law changes, the thresholding effect is compromised and a higher $\GMI$ is required. This can lead to misleading conclusions, e.g., in terms of reach prediction.

\definecolor{myc1}{HTML}{225ea8}
\definecolor{myc2}{HTML}{41b6c4}
\definecolor{myc3}{HTML}{78c679}
\definecolor{myc4}{HTML}{238443}
\definecolor{myc5}{HTML}{004529}

We conclude that performance predictors based on $\GMI$ (or variations thereof) should be used with caution. They can give rough first order estimates of the decoding performance, even if we introduce drastic changes into the channel (like, e.g., strong quantization, adding inline dispersion compensation, changing detection). Thresholds should only be used to quickly assess the performance and to determine the range of fine measurements. The accuracy is improved by modeling fairly close the channel of the system during measurement of the threshold. In all cases, actual decoding mimicking as closely as possible the true \ac{FEC} should be used.

\vspace*{-0.5ex}
\section{Performance Prediction Thresholds}

In this section, we briefly discuss some of the thresholds that are commonly used today. 

\emph{Pre-FEC SER}: The pre-FEC \ac{SER} metric is easy to compute directly from the experimental measurement. After symbol decision (e.g., using nearest-neighbor), the pre-FER SER is estimated as the fraction of wrongly decided modulation symbols. The Pre-FEC \ac{SER} should only be used as a threshold when symbol-wise \ac{HDD} is used and the constellation size is matched to the symbol size of the code. In this case, the achievable rate $\msR_\mathsf{HDD-SW}$ given by Eq.~(4) in\cite{Sheikh} is directly linked to the \ac{SER}.

\emph{Pre-FEC BER}: The easiest method to compute the pre-FEC BER is to start from estimated symbols and to use the inverse $\Phi_{\mathsf{h}}^{-1}$ of the binary labeling function $\Phi$ to generate bit sequences,  \textit{implicitly} assuming \ac{BICM} as \ac{CM} technique. The pre-FEC BER is the fraction of wrongly estimated bits after generating bit sequences for both transmit and receive symbols.
The pre-FEC BER should \textit{only} be used if bit-wise \ac{HDD} is utilized. The pre-FEC BER threshold can be used in some circumstances with \ac{SDD}, but only if the FEC that shall be evaluated has been thoroughly simulated with a model that is sufficiently close to the experimental setup (e.g., using the same modulation format, quantization, fiber model, neighboring channel setup, etc.). In that case, the use of pre-FEC BER can be tolerated.

\emph{GMI}:
Recently, the use of \ac{GMI} has become ubiquitous. We would like to point out that the notion of \ac{GMI} is a much broader concept, introduced as bound on the achievable rate for mismatched decoding\cite{Gan00}. In the case of \ac{BICM}, the \ac{GMI} equals the achievable rate. Hence, the term \ac{GMI} is often used  interchangeably with the achievable rate of \ac{BICM}. For computing the GMI, we refer to\cite{Alvarado2015b_JLT}. The GMI should \textit{only} be used for \ac{FEC} with bit-wise \ac{SDD} (i.e., for~\ac{BICM}) and if there is sufficient evidence that the code is approximately \textit{universal}.

\emph{NGMI}:
Recently, the use of probabilistic amplitude shaping become popular in optical communications\cite{buchali2016shaping}. In this case, the GMI cannot be directly used\cite{boechererAR}. A performance metric that works well in this case is the ``NGMI'' as defined in\cite{NGMI}. 

\emph{MI}:
In contrast to the GMI, the \ac{MI} is computed on symbol-level and requires the availability of the channel PDF\cite{schmalen2017performance}. 
The MI should only be used when \ac{SDD} with non-binary codes, e.g., nonbinary LDPC codes, matched to the constellation size are used as \ac{CM} scheme. The MI can also be used when multilevel coding with multi-stage decoding is employed. In many cases, the MI and GMI are relatively close (square constellations with Gray coding). In these cases, the GMI threshold can be replaced by the MI without loosing much accuracy. This has advantages as the MI is often easier to compute\cite{idler2017field}.

\vspace*{-0.7ex}
\section{Implementing Actual Decoding}\label{sec:s5_implementation}
\definecolor{mycolor0}{rgb}{0.000000,0.447000,0.741000}
\definecolor{mycolor1}{rgb}{0.850000,0.325000,0.098000}
\definecolor{mycolor2}{rgb}{0.929000,0.694000,0.125000}
\begin{figure*}[tbh!]
\includegraphics[width=\textwidth]{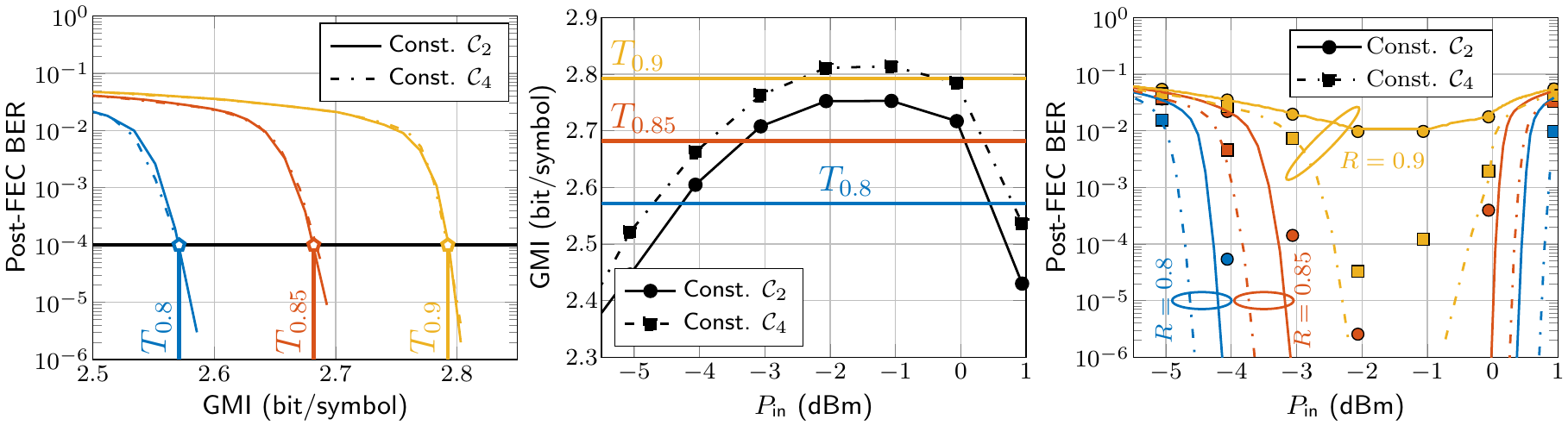}
\caption{AWGN simulation results of regular LDPC codes ($R\in\{0.8,0.85,0.9\}$), layered scaled min-sum decoding, $10$ iterations (left). Estimated GMI from experiment (middle) and performance after actual decoding and interpolation (right).}
\label{fig:simex}
\end{figure*}

Due to the pitfalls of thresholds, especially when unsure about universality, we suggest to include \ac{FEC} into the transmission experiments. We suggested in\cite{schmalen2012generic} to reuse a database of measurements to evaluate multiple \ac{FEC} schemes. We assume that the transmission experiment yields a measurement consisting of $N_M$ aligned data points $(x_\kappa, y_\kappa)$ that are given as the original complex transmit sequence $x_\kappa\in\mathbb{C}$ and the corresponding received complex sequence $y_\kappa\in\mathbb{C}$. Our method is based on the fact that the performance of most practical \ac{FEC} codes and decoders (e.g., LDPC codes) does not depend on the codeword. Hence, we assume transmission of a codeword $\bm{c}$ typically used, e.g., one that leads to the symbols following the distribution of probabilistic shaping.

\begin{figure}[t!]
\centering
\includegraphics[width=0.9\columnwidth]{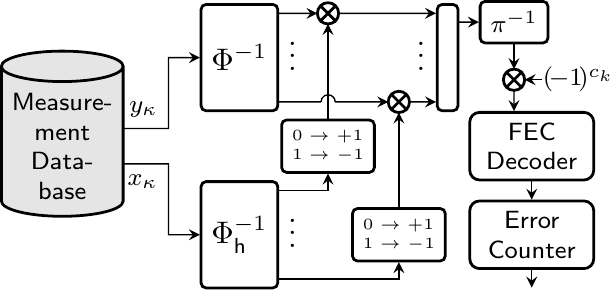}
\caption{Evaluating FEC performance from measurements for a BICM-based CM system.\vspace*{-1ex}}
\label{fig:s5_fec_evaluation_bicm}
\end{figure}

The method is illustrated in Fig.~\ref{fig:s5_fec_evaluation_bicm} for the BICM case, also used in PAS and can be easily extended to other \ac{CM} schemes. The first step of the method consists in generating an equivalent bit-stream of length $m N_M$ corresponding to the transmit sequence from the experimental database by using the inverse bit mapping function $\Phi_{\textsf{h}}^{-1}$. Similarly, the \ac{BMD} $\Phi^{-1}$ computes a sequence of \acp{LLR}. Using both sequences, we generate a set of equivalent \acp{LLR}, corresponding to the transmission of the selected codeword $\bm{c}$ which can be fed to an \ac{FEC} decoder.

\vspace*{-0.5ex}
\section{Performance Example}\label{sec:s5_performance_example}

To illustrate the proposed method, we consider a system using 8QAM together with \ac{BICM}. We reuse the measurement database already used in\cite{schmalen2017performance} with the two 8QAM constellations denoted $\mathcal{C}_2$ and $\mathcal{C}_4$  in\cite{schmalen2017performance} together with their \ac{GMI}-maximizing bit labeling. We use 3 regular QC-\ac{LDPC} codes with rates $R\in\{0.8,0.85,0.9\}$ of length $n=38400$\,bit. Decoding takes place using $I=10$ iterations with a layered scaled min-sum decoder ($\alpha=0.75$). The AWGN performance of the codes is shown in Fig.~\ref{fig:simex} (left) as a function of the GMI.  We see that the codes act approximately universal and their performance only marginally depends on the chosen constellation. For a post-FEC BER of $10^{-4}$ , we find equivalent achievable rate thresholds $T_R$.

Transmission takes place over a coherent, dual-polarization fiber-optic communication system at a symbol rate of $41.6$~Gbaud over 8 round trips in a re-circulating loop ($3200$\,km), described in\cite{schmalen2017performance}. The transmission test-bed consists of one laser under test and additionally 63 loading channels spaced by 50\,GHz.  Fig.~\ref{fig:simex} (middle) shows the estimated GMI as function of the input power $P_{\text{in}}$ per \ac{WDM} channel with the thresholds $T_R$ for $R\in\{0.8,0.85,0.9\}$. Whenever the estimated achievable rate lies above the threshold $T_R$, successful transmission is possible (i.e., post-\ac{FEC} \ac{BER} below $10^{-4}$). For example, consider $T_{0.9}$:  with constellation $\mathcal{C}_4$, we are just barely above the line for $P_{\text{in}}\in\{-2\,\text{dBm},-1\,\text{dBm}\}$, which means that decoding is also only barely possible. Contrary, with  $\mathcal{C}_2$, reliable decoding is not possible.

In Fig.~\ref{fig:simex} (right), we use the AWGN simulation to estimate the post-FEC BER of the transmission system by interpolation (solid and dash-dotted lines). Additionally, we carried out actual decoding using the \ac{LDPC} codes (solid markers). We can see that the estimates from interpolation match the actual decoding performance closely, especially for high BERs. However, we see another deviation at low BERs which are caused mostly by non-stationarity in the measurements.\vspace*{-1ex}


\bibliographystyle{abbrv}

\begin{thebibliography}{11}
\vspace{2pt}
\scriptsize
\bibitem{schmalen2013bltj}
L.~Schmalen et al., ``Forward error correction in optical core and optical access networks,'' \textit{Bell Labs Technical Journal} (2013).
\bibitem{LevenMI}
A.~Leven et al., ``Estimation
  of soft {FEC} performance in optical transmission experiments,'' \textit{{IEEE}
  Photon. Technol. Lett.} (2011).
\bibitem{Alvarado2015b_JLT}
A.~Alvarado et al., ``Replacing the
  soft-decision {FEC} limit paradigm in the design of optical communication
  systems,'' \textit{J. Lightw. Technol.}, (2015).
\bibitem{Franceschini06}
M.~Franceschini et al., ``Does the performance of {LDPC}
  codes depend on the channel?'' \textit{{IEEE} Trans. Commun.} (2006).
\bibitem{sanaei2008design}
A.~Sanaei et al., ``On the design of universal {LDPC}
  codes,'' in \textit{Proc. ISIT} (2008).
\bibitem{schmalen2017performance}
L.~Schmalen et al., ``Performance prediction of
  nonbinary forward error correction in optical transmission experiments,''
  \textit{J. Lightw. Technol.} (2017).
\bibitem{Sheikh}
A. Sheikh et al., ``Achievable information rates for coded modulation with hard decision decoding for coherent fiber-optic systems,'' \textit{J. Lightw. Technol.} (2017)
\bibitem{Gan00}
A.~Ganti et al., ``Mismatched decoding revisited:
  general alphabets, channels with memory, and the wide-band limit,''
  \textit{IEEE Trans. Inf. Theory} (2000).
\bibitem{buchali2016shaping}
F.~Buchali et al., ``Rate adaptation and reach increase by probabilistically shaped {64-QAM}: An
  experimental demonstration,'' \textit{J. Lightw. Technol.} (2016).
\bibitem{boechererAR}
G. B\"ocherer, ``Achievable rates for probabilistic shaping,'' arXiV:1707:01134
\bibitem{NGMI}
J. Cho et al., ``Normalized generalized mutual information as a forward error correction threshold for probabilistically shaped QAM,'' in \textit{Proc. ECOC} (2017).
\bibitem{idler2017field}
W.~Idler et al. ``Field trial of a {1 Tb/s} super-channel network
  using probabilistically shaped constellations,'' \textit{J. Lightw. Technol.}
  (2017).
\bibitem{schmalen2012generic}
L.~Schmalen et al., ``A generic tool for
  assessing the {soft-FEC} performance in optical transmission experiments,''
  \textit{{IEEE} Photon. Technol. Lett.} (2012).\end{thebibliography}
\begin{spacing}{0.45}
\setlength{\bibsep}{1pt}

\end{spacing}
\vspace{-4mm}

\end{document}